\documentclass[preprint,12pt]{elsarticle}
\usepackage{amssymb,amsmath,amscd}
\usepackage{amsfonts,bm,latexsym}

\allowdisplaybreaks
\journal{Physics Letters B}

\def\be{\begin{equation}}
\def\ee{\end{equation}}
\def\bea{\begin{eqnarray}}
\def\eea{\end{eqnarray}}

\def\p{\partial}

\begin{document}

\begin{frontmatter}

\title{Thermodynamics of static dyonic AdS black holes in the $\omega$-deformed Kaluza-Klein
gauged supergravity theory}

\author{Shuang-Qing Wu\footnote{\textit{Email address}: sqwu@cwnu.edu.cn, Corresponding author},
Shoulong Li\footnote{\textit{Email address}: sllee.phys@gmail.com}}
\address{Institute of Theoretical Physics, China West Normal University, Nanchong, Sichuan 637002,
People's Republic of China}

\begin{abstract}
We study thermodynamical properties of static dyonic AdS black holes in four-dimensional $\omega$-deformed
Kaluza-Klein gauged supergravity theory, and find that the differential first law requires a modification
via introducing a new pair of thermodynamical conjugate variables ($X, Y$). To ensure such a modification,
we then apply the quasi-local ADT formalism developed in Ref. [20] to calculate the quasi-local conserved
charge and identify that the new pair is precisely the one previously introduced to modify the differential
form of the first law.
\end{abstract}

\begin{keyword}
black hole thermodynamics \sep dyonic AdS black hole \sep conserved charge \sep Kaluza-Klein supergravity

\PACS 04.70.Dy \sep 04.65.+e \sep 04.50.Cd
\end{keyword}

\end{frontmatter}

\section{Introduction}

Black holes are the most important compact objects predicted soon after Albert Einstein built 
his General Theory of Relativity \cite{AE1915} just one century ago. With lots of exact solutions being
found, there are many researches on various different kinds of black holes and a large amount of great
success has been achieved in the area of black hole physics. One of the remarkable achievements is the
four laws of black hole thermodynamics established by Bardeen, Carter and Hawking \cite{BCH73} via the
analogy with those of ordinary thermodynamical system. In particular, the differential first law is
expressed as $dM = T\, dS +\Omega\, dJ +\Phi\, dQ +\Psi\, dP$ for a stationary asymptotically flat black
hole. In this formula, ($M, J, Q, P$) are the mass, angular momentum, electric and magnetic charge
measured at infinity, while ($T, S, \Omega, \Phi, \Psi$) represent the temperature, entropy, angular
velocity of the horizon, electro-static and magnetic potentials at the horizon, respectively.

However, it has been demonstrated in certain cases \cite{GKK96,HM04,HMTZ0247,LPPVP12,LLP14,LPP13,CC14}
that thermodynamics might receive necessary modifications due to the presence of nontrivial matter fields.
If the asymptotic behavior of a scalar field at infinity is $\phi = \phi_\infty +\phi_1/r +\phi_2/r^2
+\cdots$ in the asymptotically flat spacetimes, it was shown in Ref. \cite{GKK96} that under variation
of moduli field $\phi_{\infty}$, the first law of black hole thermodynamics becomes $dM = T\, dS +\Omega\,
dJ +\Phi\, dQ +\Psi\, dP -\Sigma\, d\phi_{\infty}$, where $\Sigma = \phi_1$ is the scalar charge. In the
cases of asymptotically AdS spacetimes with $\phi_\infty = 0$, the contribution of a scalar field to
black hole thermodynamics was previously studied in Refs. \cite{HM04,HMTZ0247}. In recent researches
\cite{LPPVP12,LLP14,LPP13} on thermodynamics of spherically symmetric static AdS black holes, such as
the solution given in conformal gravity \cite{LPPVP12} and the one presented in Einstein-Proca theory
\cite{LLP14} as well as dyonic black holes found in Kaluza-Klein (KK) gauge supergravity theory \cite{LPP13},
the first law should be modified by extra hairs. It was found that in Ref. \cite{LPPVP12}, the modification
of the first law is due to the spin-2 hair, while in Ref. \cite{LLP14} the contribution comes from a massive
spin-1 modes, rather than the massive spin-2 modes. In four-dimensional asymptotically AdS spacetimes, it
was shown that a massless scalar with the large-$r$ boundary behavior: $\phi = \phi_1/r +\phi_2/r^2 +\cdots$
can break some of the boundary AdS symmetries unless one of the following three conditions is satisfied:
(1) $\phi_1 = 0$, or (2) $\phi_2 = 0$, or (3) $\phi_2/\phi_1^2$ is a fixed constant \cite{HM04}. (See also
Refs. \cite{HMTZ0247} and \cite{LL14}.) However, it should be pointed out that if the mass squared of
the four-dimensional scalar field is -2, the aforementioned cases are the ones preserving the full SO(2,3)
symmetry.

In this Letter, we will focus on the case where the modification of the first law is due to the scalar hair
of static dyonic AdS black holes \cite{LPP13,CC14} in ($\omega$-deformed) KK gauged supergravity theory. In
Ref. \cite{LPP13}, a static dyonic AdS black hole solution in four-dimensional KK gauged supergravity was
constructed, for which the scalar boundary behavior violates the above-mentioned three criteria \cite{HH05}.
It is proposed in Ref. \cite{LPP13} that the first law should be rescued via introducing a new pair of
thermodynamical conjugate variables $(X, Y)$, so that the differential first law can be rewritten as $dM =
T\, dS +\Omega\, dJ +\Phi\, dQ +\Psi\, dP -X\, dY$. On the contrary, Chow and Compere \cite{CC14} presented
general static AdS black hole solution in four-dimensional $\mathcal{N} = 2$, STU gauged supergravity and
insisted that the first law is unchanged while the mass is non-integrable due to the non-existence of a
conserved symplectic structure in covariant phase space. Obviously, it deserves a deeper investigation
of this issue, see Ref. \cite{QW15} for a recent discussion on this subject.

On the other hand, the maximal $\mathcal{N} = 8$, SO(8) gauged supergravity theory constructed in Ref.
\cite{WN82} has been regarded uniquely for a long time. But a recent evidence \cite{AIT12} demonstrates
that there is a one-parameter family of inequivalent SO(8) gauged supergravity theories characterized
by an angular parameter $\omega$. In recent years, there has been a lot of great interest to study various
different consistent truncations \cite{BGRDVBHT,CPPRBS,LPP14} of the $\omega$-deformed maximal $\mathcal{N}
= 8$, SO(8) gauged supergravity, especially the truncations of scalar fields. Some black hole solutions
in $\omega$-deformed gauged $\mathcal{N} = 8$ supergravity was constructed recently in Ref. \cite{AA14}.
What most interested us here is the truncation to the $\omega$-deformed KK gauged supergravity theory
and exact solutions to it \cite{LPP14}.

The aim of this Letter is to investigate thermodynamical properties of spherically symmetric static dyonic
AdS black holes in four-dimensional $\omega$-deformed KK gauged supergravity theory \cite{LPP14}. Similar
to the analysis done in the un-deformed case \cite{LPP13}, we find that the differential first law needs
a modification by introducing a new pair of thermodynamical conjugate variables ($X, Y$). To ensure such
a modification, it is necessary to calculate the conserved charge via another different approach. Compared
with the Wald's formalism \cite{W93IW94} adopted in Ref. \cite{LPP13}, in this work we will apply the
quasi-local ADT formalism \cite{KKY13} to derive the conserved charge and deduce that the new conjugate
pair ($X, Y$) is precisely that previously introduced \cite{LPP13} to modify the differential form of
the first law.

The remaining part of this Letter is organized as follows. In Section 2, we study the thermodynamics
of static dyonic AdS black holes in the $\omega$-deformed KK gauged supergravity theory. We check the
differential first law and the Bekenstein-Smarr formula. Inferred from the modification given in \cite{LPP13},
we obtain precisely the same modification of the differential first law. In Section 3, the conserved
charge is calculated by using the quasi-local ADT formalism to identify the modification with the
quantity introduced in Section 2. Finally, we present our conclusions with some comments.

\section{Thermodynamics of static dyonic AdS black
holes in $\omega$-deformed KK gauged supergravity}

In Ref. \cite{LPP14}, the $\omega$-deformed KK gauged supergravity theory is obtained via a series of
consistent truncations of the $\omega$-deformed maximal $\mathcal{N} = 8$, SO(8) gauged supergravity.
The Lagrangian for this theory is given below:
\be\begin{split}
e^{-1}\mathcal{L} &= R +6g^2\cosh\phi -\frac{3}{2}(\p\phi)^2 \\
&\quad -\frac{2F_{\mu\nu}F^{\mu\nu} +\sin{2\omega}\sinh{3\phi}\epsilon^{\mu\nu\rho\sigma}
 F_{\mu\nu}F_{\rho\sigma}}{8(e^{3\phi}\cos^2\omega +e^{-3\phi}\sin^2\omega)} \, ,
\end{split}\ee
in which $g$ is the cosmological constant and $\omega$ is a deformed parameter. Note that in the un-deformed
case where $\omega = 0$, the above theory returns to the KK gauged supergravity theory. If we set $g = 0$
further, then it reduces to the standard KK supergravity theory obtained from the $S^1$ reduction of the
five-dimensional pure Einstein's gravity theory.

A four-dimensional static dyonic AdS black hole solution is also presented there \cite{LPP14}, for which
the metric, the dilaton scalar, the U(1) gauge potential and its dual are given below
\begin{gather}
ds^2 = \sqrt{H_1(r)H_2(r)}\Big[-\frac{f(r)\, dt^2}{H_1(r)H_2(r)} +\frac{dr^2}{f(r)} +r^2(d\theta^2
 +\sin^2\theta\, d\varphi^2)\Big] \, , \\
\phi = \frac{1}{2}\ln\Big[\frac{H_2(r)}{H_1(r)}\Big] \, , \qquad\quad
f(r) = 1 -\frac{2m}{r} +g^2r^2H_1(r)H_2(r) \, , \\
A = \Big[\frac{h_1(r)}{H_1(r)}\cos\omega -\frac{h_2(r)}{H_2(r)}\sin\omega \Big]\, dt
 +4(P\cos\omega +Q\sin\omega)\cos\theta\, d\varphi \, , \\
\widetilde{A} = \Big[\frac{h_1(r)}{H_1(r)}\sin\omega +\frac{h_2(r)}{H_2(r)}\cos\omega \Big]\, dt
 +4(P\sin\omega -Q\cos\omega)\cos\theta\, d\varphi \, ,
\end{gather}
where
\be\begin{split}
h_1(r) &= \frac{4Q}{q}\Big(\frac{p}{r} +\frac{p+q}{q-2m}\Big) \, , \quad
h_2(r) = \frac{4P}{p}\Big(\frac{q}{r} +\frac{p+q}{p-2m}\Big) \, , \\
H_1(r) &= 1 +\frac{q-2m}{r} +\frac{q(p-2m)(q-2m)}{2(p+q)r^2} \, ,  \\
H_2(r) &= 1 +\frac{p-2m}{r} +\frac{p(p-2m)(q-2m)}{2(p+q)r^2} \, ,
\end{split}\ee
in which
\be
P = \frac{\sqrt{p(p^2-4m^2)}}{4\sqrt{p+q}} \, , \quad Q = \frac{\sqrt{q(q^2-4m^2)}}{4\sqrt{p+q}} \, ,
\ee
are electric and magnetic charges of the static dyonic AdS black hole in the un-deformed KK gauged
supergravity theory \cite{LPP13}, expressed in terms of the parameters $p\geq 2m$ and $q\geq 2m$.
The relation between ($p, q$) and $(\beta_1, \beta_2$) used in Ref. \cite{LPP13} is
\be
\beta_1 = \frac{p(q-2m)}{q(p+2m)} \, , \quad \beta_2 = \frac{q(p-2m)}{p(q+2m)} \, .
\label{pares}
\ee
Compared with the solution given in Ref. \cite{LPP13}, the only modification is the U(1) gauge potentials
obtained via a duality rotation of those in \cite{LPP13}, while the metric and the dilaton scalar remain
unchanged. If we set $\omega = 0$ and $p = q$, then the dilaton vanishes and the solution recovers to the
dyonic Reissner-Nordstr\"{o}m AdS black hole in four-dimensional Einstein-Maxwell theory.

The event horizon is defined through $f(r_+) = 0$. The temperature and entropy of the horizon are easily
calculated as
\be
T = \frac{f^\prime(r_+)}{4\pi\sqrt{H_1(r_+)H_2(r_+)}} \, , \quad
S = \pi r_+^2\sqrt{H_1(r_+)H_2(r_+)} \, .
\ee

The electro-static and magnetic potentials are given by
\be\begin{split}
\Phi_{\omega} &= A_t\big|_{r_+} -A_t\big|_{\infty} = \Phi\cos\omega -\Psi\sin\omega \, , \\
\Psi_{\omega} &= \widetilde{A}_t\big|_{r_+} -\widetilde{A}_t\big|_{\infty}
 = \Phi\sin\omega +\Psi\cos\omega \, ,
\end{split}\ee
where
\be
\Phi = \frac{h_1(r_+)}{H_1(r_+)} -\frac{\sqrt{(q+2m)(p+q)}}{\sqrt{q(q-2m)}} \, , \quad
\Psi = \frac{h_2(r_+)}{H_2(r_+)} -\frac{\sqrt{(p+2m)(p+q)}}{\sqrt{p(p-2m)}} \, .
\ee
The electric and magnetic charges can be computed as
\be
Q_{\omega} = Q\cos\omega -P\sin\omega \, , \quad P_{\omega} = Q\sin\omega +P\cos\omega \, .
\ee
Note that the above expressions are expressed in terms of their counterparts in the un-deformed case
\cite{LPP13} after considering the relation (\ref{pares}).

Using the conformal Weyl tensor method \cite{KCLPW}, it is not difficult to calculate the mass
\be
M = \frac{p+q}{4} \, . \label{mass}
\ee

Now let's check the differential first law of thermodynamics. The first law does hold for $g = 0$,
and is written as $dM = T\, dS +\Phi_{\omega}\, dQ_{\omega} +\Psi_{\omega}\, dP_{\omega}$. For $g\ne 0$,
it is apparent that the first law still holds true in the cases: (1) $p = 2m$, or (2) $q = 2m$, or (3)
$p = q$ if $g$ is not viewed as a thermodynamical viable. In other case, the differential first law
no longer holds true and must be compensated via introducing a new conjugate pair ($X, Y$)
\be
dM = T\, dS +\Phi_{\omega}\, dQ_{\omega} +\Psi_{\omega}\, dP_{\omega} -X\, dY \, ,
\ee
where
\be
X = g^2\frac{(p-q)(p^2-4m^2)^{3/2}(q^2-4m^2)^{1/2}}{16(p+q)^2} \, , \quad
Y = \sqrt{\frac{q^2-4m^2}{p^2-4m^2}} \, ,
\ee
are the same expressions as those given in Ref. \cite{LPP13}. This is easily verified by using (\ref{pares}),
and it can be proved that
\be\begin{split}
-X\, dY &= d M -T\, dS -\Phi_{\omega}\, dQ_{\omega} -\Phi_{\omega}\, dP_{\omega} \\
&= dM -T\, dS -\Phi\, dQ -\Psi\, dP \, ,
\end{split}\ee
which means that $\omega$ doesn't change the first law of thermodynamics.

One can further treat the cosmological constant as a generalized ``pressure" $\mathcal{P} = 3g^2/(8\pi)$,
its conjugate quantity $\mathcal{V}$ as a thermodynamical volume \cite{CGKP11}, then the differential first
law reads
\be\begin{split}
dM &= T\, dS +\Phi_{\omega}\, dQ_{\omega} +\Psi_{\omega}\, dP_{\omega}
 -X\, dY +\mathcal{V}\, d\mathcal{P} \\
 &= T\, dS +\Phi\, dQ +\Psi\, dP -X\, dY +\mathcal{V}\, d\mathcal{P} \, ,
\end{split}\ee
where
\be
\mathcal{V} = \frac{4\pi}{3}\Big[r_+^3 +\frac{3}{4}(p+q-4m)r_+^2
 +\frac{1}{4}(p-2m)(q-2m)\Big(3r_+ -m +\frac{pq}{p+q}\Big)\Big] \, .
\ee
One can also verify that the Bekenstein-Smarr formula is given by
\be\begin{split}
M &= 2T\, S +\Phi_{\omega}\, Q_{\omega} +\Psi_{\omega}\, P_{\omega} -2\mathcal{V}\, \mathcal{P} \\
& = 2T\, S +\Phi\, Q +\Psi\, P -2\mathcal{V}\, \mathcal{P} \, ,
\end{split}\ee
and is independent of the deformation parameter $\omega$. Moreover, the ($X, Y$) pair doesn't appear
in the integral first law.

\section{Quasi-local conserved charge}

In the last section, we have approved the viewpoint proposed in Ref. \cite{LPP13} and followed the same
recipe to modify the differential first law by introducing a new pair of thermodynamical variables $(X, Y)$.
However, the precise physics origin of $(X, Y)$ still remains a mystery. In order to make an in-depth
analysis of the $(X, Y)$ pair, we shall adopt the quasi-local ADT formalism \cite{KKY13} rather than
the covariant phase space approach \cite{W93IW94} used in \cite{LPP13} to calculate the conserved charge
and study the fall-off behavior of the scalar field at infinity.

As far as the conserved charge of AdS black hole is concerned, up to date there are many different methods
available to calculate it, such as the covariant phase space approach \cite{W93IW94}, cohomological method
\cite{BBCMN}, Ashtekar-Magnon-Das formalism \cite{AMAD}, Abbott-Deser-Tekin (ADT) formalism \cite{ADDT},
and quasi-local ADT formalism \cite{KKY13}. For an earlier review on the quasi-local conserved charge,
see Ref. \cite{LBS09}. The quasi-local ADT formalism \cite{KKY13} is a novel way to calculate the conserved
charge at finite spacetime domains and receives a lot of recent attention \cite{HPYHJPY145,JP14}. Speaking
roughly, one can establish a one-to-one correspondence between the ADT potential and the off-shell linear
Noether potential by considering an appropriate variation of metric. From the off-shell ADT potential, one
can easily construct the quasi-local charge. Note that the same result can be arrived at by varying Bianchi
identity \cite{JP14}.

Taking the variation of the action, we get
\be
\delta S = \frac{1}{16\pi}\int d^4x \sqrt{-g}\big(E_{\Psi} \delta\Psi +\nabla_\mu\Theta^\mu\big) \, ,
\ee
where $E_{\Psi} \delta\Psi = E_{(g)\mu\nu}\delta g^{\mu\nu} +E_{\phi}\delta\phi +E_{A}^\nu\delta A_\nu$
and $\Theta^\mu = \Theta_{(g)}^\mu +\Theta_\phi^\mu +\Theta_A^\mu$ denote the equation of motion and the
surface term, respectively. Considering the infinitesimal diffeomorphism: $x^\mu \to x^\mu +\xi^\mu$,
one can deduce the off-shell Noether current $\mathcal{J}^\mu$ through equating the diffeomorphism to the
general variation as
\be
\mathcal{J}^\mu = 2\mathcal{E}^{\mu\nu}\xi_\nu +\xi^\mu\mathcal{L} -\Theta^\mu \, ,
\ee
where we have used the off-shell identity: $2\xi_\nu\nabla_\mu E_{(g)}^{\mu\nu} +E_{\phi} \delta_\xi\phi
+E_{A} \delta_\xi A = \nabla_\mu(Z^{\mu\nu}\xi_\nu)$, and denoted $\mathcal{E}^{\mu\nu} = E_{(g)}^{\mu\nu}
-Z^{\mu\nu}/2$. Then one can introduce the off-shell Noether potential $K^{\mu\nu}$ by using $\mathcal{J}^\mu
= \nabla_\nu K^{\mu\nu}$, in which
\be
K^{\mu\nu} = 2\nabla^{[\mu}\xi^{\nu]} +\big[k(\phi)F^{\mu\nu}
 -4\hat{k}(\phi)\epsilon^{\mu\nu\rho\sigma}F_{\rho\sigma}\big] A_\lambda\xi^\lambda \, ,
\ee
where
$$
k(\phi) = \frac{1}{e^{3\phi}\cos^2\omega +e^{-3\phi}\sin^2\omega} \, , \quad
 \hat{k}(\phi) = \frac{-\sin{2\omega}\sinh{3\phi}}{2(e^{3\phi}\cos^2\omega +e^{-3\phi}\sin^2\omega)} \, .
$$

Now let's define the ADT current $\mathcal{J}_{ADT}^\mu$ which reads
\be
\mathcal{J}_{ADT}^\mu = \xi_\nu\delta\mathcal{E}^{\mu\nu}
 +\frac{1}{2}g^{\alpha\beta}\mathcal{E}^{\mu\nu}\xi_\nu\delta g_{\alpha\beta}
 +\mathcal{E}^{\mu\nu}\xi^\rho\delta g_{\nu\rho} +\frac{1}{2} \xi^\mu E_{\Psi} \delta\Psi \, ,
\ee
and write it in a compact form: $\sqrt{-g}\mathcal{J}_{ADT}^\mu = \delta(\sqrt{-g}\mathcal{E}^{\mu\nu}\xi_\nu)
+\sqrt{-g}\xi^\mu E_{\Psi} \delta\Psi/2$. The corresponding off-shell ADT potential $\mathcal{Q}_{ADT}^{\mu\nu}$
is introduced by $\mathcal{J}_{ADT}^\mu = \nabla_\nu\mathcal{Q}_{ADT}^{\mu\nu}$. To find out the relationship
between the off-shell Noether current for the infinitesimal diffeomorphism and the linearized conserved current
for a Killing vector, we now take $\xi^\mu = (\p_t)^\mu$ and consider the change in the Noether potential
$K^{\mu\nu}$, then we can get
\be
\mathcal{Q}_{ADT}^{\mu\nu} = \frac{1}{2} \delta K^{\mu\nu}
 +\frac{1}{4}K^{\mu\nu}g^{\alpha\beta}\delta g_{\alpha\beta} -\xi^{[\mu}\Theta^{\nu]} \, ,
\ee
from which the off-shell ADT potential $\mathcal{Q}_{ADT}^{\mu\nu}$ is rewritten as \cite{BBCMN}
\be
\mathcal{Q}_{ADT}^{\mu\nu} = \mathcal{Q}_{(g)}^{\mu\nu} +\mathcal{Q}_{\phi}^{\mu\nu}
 +\mathcal{Q}_{F}^{\mu\nu} +\mathcal{Q}_{CS}^{\mu\nu} \, ,
\ee
where
\begin{subequations}
\begin{alignat}{1}
\mathcal{Q}_{(g)}^{\mu\nu} &= \frac{1}{2} h\nabla^{[\mu}\xi^{\nu]} +\xi^{[\mu}\nabla^{\nu]}h
 +\xi_\alpha\nabla^{[\mu} h^{\nu]\alpha} -h^{\alpha[\mu}\nabla_\alpha\xi^{\nu]}
 -\xi^{[\mu}\nabla_\alpha h^{\nu]\alpha} \, , \\
\begin{split}
\mathcal{Q}_{F}^{\mu\nu} &= \frac{1}{2}\Big[F^{\mu\nu}\frac{\p k(\phi)}{\p\phi^a}\delta\phi^a
 +k(\phi)\delta F^{\mu\nu} +\frac{h}{2}k(\phi)F^{\mu\nu} -2k(\phi)h^{\mu\lambda}F_\lambda^{\nu}\Big] \\
&\quad \times (A_\lambda\xi^\lambda +Cst) +k(\phi)\big(\xi^\lambda F^{\mu\nu}
 +\xi^{[\mu}\nabla_\alpha F^{\nu]\alpha}\big)\delta A_\lambda \, ,
\end{split} \\
\begin{split}
\mathcal{Q}_{CS}^{\mu\nu} &= -2\epsilon^{\mu\nu\rho\sigma}
 \Big[F_{\rho\sigma}\frac{\p\hat{k}(\phi)}{\p\phi^a}\delta\phi^a
 +\hat{k}(\phi)\delta F_{\rho\sigma}\Big](A_\lambda\xi^\lambda +Cst) \\
&\quad -2\big[\epsilon^{\mu\nu\rho\sigma}\xi^\lambda +2\xi^{[\mu}\epsilon^{\nu]
 \lambda\rho\sigma}\big]\hat{k}(\phi)F_{\rho\sigma}\delta A_\lambda \, ,
\end{split} \\
\mathcal{Q}_{\phi}^{\mu\nu} &= 3\xi^{[\mu}\nabla^{\nu]}\phi\, \delta\phi \, ,
\end{alignat}
\end{subequations}
in which $h_{\mu\nu} = \delta g_{\mu\nu}$, $h^{\mu\nu} = -\delta g^{\mu\nu}$, $h = g^{\mu\nu}\delta
g_{\mu\nu}$, $\delta F^{\mu\nu} = g^{\mu\alpha}g^{\nu\beta}\delta F_{\alpha\beta}$.

One can choose the gauge $\delta A_\mu = \mathcal{L}_\xi A_\mu = 0$ so that $d(A_\mu\xi^\mu)$ is gauge
invariant. For our aim to calculate the conserved mass, a convenient choice for the gauge constant $Cst$
that makes $(A_\mu\xi^\mu +Cst)|_\infty = 0$ yields
\be
Cst = \frac{\sqrt{(q+2m)(p+q)}}{\sqrt{q(q-2m)}}\cos\omega
-\frac{\sqrt{(p+2m)(p+q)}}{\sqrt{p(p-2m)}}\sin\omega \, , \label{gauge}
\ee
which is equivalent to making a gauge transformation on $A_{\mu}$ so that the electro-static potential
or the $t$-component of the shifted potential vanishes at infinity.

According to Ref. \cite{BBCMN}, for a class of one-parameter path in the solution space, one can define
a path-independent quasi-local conserved charge as
\be
\mathcal{Q} = \frac{1}{8\pi}\int \sqrt{-g}\mathcal{Q}_{ADT}^{\mu\nu}d\Sigma_{\mu\nu} \, .
\ee

For our final aim to obtain the conserved charge, we can introduce a new radial coordinate $\rho$ which
is convenient for us to study the asymptotic fall-off behaviors of the metric and the matter fields.
Then the line element becomes
\be
ds^2 = -f(\rho)dt^2 +\frac{d\rho^2}{h(\rho)f(\rho)} +\rho^2(d\theta^2 +\sin^2\theta\, d\varphi^2) \, .
\ee
For large $\rho$, we get
\be\begin{split}
f(\rho) &= g^2\rho^2 +1 -\frac{p+q}{2\rho} +\frac{p^2+q^2 -pq-4m^2}{4\rho^2} +\cdots \, , \\
h(\rho) &= 1 +\frac{3(p-q)^2}{16\rho^2} +\frac{(p-q)^2(8m^2-p^2-q^2)}{8(p+q)\rho^3} +\cdots \, , \\
\phi &= \frac{p-q}{2\rho} +\frac{(p-q)(8m^2-p^2-q^2)}{8(p+q)\rho^2} +\cdots \, ,
\end{split}\ee
and choose an infinitesimal parametrization of a one-parameter path in the solution space by letting
\be
m \to m +dm \, , \quad p \to p +dp \, , \quad q \to q +dq \, .
\ee
After some tedious algebra manipulations and using the gauge choice (\ref{gauge}), we can obtain the
infinitesimal charge
\be\begin{split}
d\mathcal{Q} &= \frac{1}{4}(dp +dq) -\frac{g^2(p-q)^2}{4(p+q)}m\, dm \\
&\quad +\frac{g^2(p-q)}{16(p+q)^2}\big[(p^2-4m^2)q\, dq -(q^2-4m^2)p\, dp\big] \, .
\end{split}\ee
Using Eq. (\ref{mass}), we finally get
\be\begin{split}
X\, dY &= d\mathcal{Q} -dM \\
& = \frac{g^2(p-q)}{32(p+q)^2}\big[(p^2-4m^2)\, d(q^2-4m^2) -(q^2-4m^2)\, d(p^2-4m^2)\big] \, .
\end{split}\ee
In the general cases, $\mathcal{Q}$ is not integrable because $d^2\mathcal{Q} = dX\wedge dY \not= 0$ unless
one of the following conditions is satisfied: (1) $g = 0$, or (2) $p = 2m$, or (3) $q = 2m$, or (4) $p = q$,
or (5) $p^2-4m^2 = c(q^2-4m^2)$ for an arbitrary constant $c$. Note that in the above analysis, we have treated
the cosmological constant $g$ as a true constant ($dg = 0$). If $g$ is viewed as a variable, then $\mathcal{Q}$
is still not integrable and one must add a counter-term $\mathcal{V}\, d\mathcal{P}$ to cancel the corresponding
divergent term in the expression of $d\mathcal{Q}$. Therefore it is reasonable to infer that the differential
first law of thermodynamics needs a modification via the $(X, Y)$ pair.

At last, we would like to check that $X\, dY $ is the exact modification given before. According to the analysis
made in Ref. \cite{LPP13}, we have
\be
h(\rho)f(\rho) -f(\rho) = \frac{1}{4}g^2\phi_1^2 +\frac{2}{3\rho}g^2\phi_1\phi_2 +\cdots \, ,
\quad \phi = \frac{\phi_1}\rho +\frac{\phi_2}{\rho^2} +\cdots \, ,
\ee
where
\be
\phi_1 = \frac{p-q}{2} \, , \quad \phi_2 = \frac{(p-q)(8m^2-p^2-q^2)}{8(p+q)} \, ,
\ee
so we can re-express \cite{LPP13}
\be
X\, dY = \frac{1}{12}g^2(2\phi_2\delta \phi_1 -\phi_1\delta \phi_2) \, ,
\ee
which vanishes when (1) $\phi_1 = 0$, or (2) $\phi_2 = 0$, or (3) $\phi_2 = c\phi_1^2$.

\section{Conclusion}

In this Letter, we have studied thermodynamical properties of static dyonic AdS black hole in four-dimensional
$\omega$-deformed KK gauged theory, and obtained the expression of the modified term for the differential first
law. Although the similar problem in the un-deformed theory has been tackled previously in Ref. \cite{LPP13} by
using the Wald's procedure, here we have adopted a different method based on the quasi-local ADT formalism to
compute the conserved charge. We approve the proposal in Ref. \cite{LPP13} to modify the differential first law
by adding a new term $X\, dY$. Furthermore, we find that the deformation parameter $\omega$ doesn't change the
first law.

Although the differential first law can be satisfactorily restored via the introduction of the ($X, Y$) pair,
this is somehow by heart. Do these two quantities have a general universal definition, and what is the genuine
physics hidden behind them?

\section*{Acknowledgements}
S.Q. Wu is supported by the National Natural Science Foundation of China (NSFC) under Grant No. 11275157.
S.L. Li is grateful to Dr. J.J. Peng for helpful discussions and our referee for good advice to improve
the presentation of this work.

\section*{Reference}

\def\JHEP{J. High Energy Phys.~}

\end{document}